      [1999/12/01 v1.4c Il Nuovo Cimento]
\documentclass{cimento}

\usepackage{graphicx}
\title{Optically Selected GRB Afterglows, a Real Time Analysis System at the CFHT}
\author{F.~Malacrino\from{ins:omp}\ETC,
J-L.~Atteia\from{ins:omp},
M.~Boer\from{ins:ohp},
A.~Klotz\from{ins:omp}\\
JJ.~Kavelaars\from{ins:can}
        \atque
J-C.~Cuillandre\from{ins:haw}}
\instlist{\inst{ins:omp} Observatoire Midi-Pyr\'en\'ees - Toulouse,
       France
  \inst{ins:ohp} Observatoire de Haute-Provence, St Michel, France
  \inst{ins:can} Herzberg Institute for Astrophysics, CADC, Canada
  \inst{ins:haw} Canada France Hawaii Telescope, Hawaii, US
    }
\begin{document}

\maketitle

\begin{abstract}
We attempt to detect optical GRB afterglows on images taken by the Canada France Hawaii Telescope 
for the Very Wide survey, component of the Legacy Survey. To do so, a Real Time Analysis System 
called "Optically Selected GRB Afterglows" has been installed on a dedicated computer in Hawaii. 
This pipeline automatically and quickly analyzes Megacam images and extracts from them a list of 
variable objects which is displayed on a web page for validation by a member of the collaboration.
The Very Wide survey covers 1200 square degrees down to i'=23.5.
This paper briefly explain the RTAS process.



\end{abstract}

\section{The Canada France Hawaii Telescope and its Legacy Survey}

\subsection{CHFT at glance}
The CFHT is a 3.6 m telescope located on the Mauna Kea in Hawaii. 
Built in the late 70's, it has been recently equiped with a new instrument, Megacam. 
The specifications of this CCD imager are the following:
\begin{itemize}
\item[-] 36 2048x4612 pixel CCDs with 0.185 arcsec/pixel resolution
\item[-] 1 square degree field of view
\item[-] 5 filters set (u*, g', r', i', z')
\end{itemize}

\begin{figure}[h]
\begin{center}
\includegraphics[height=.075\textheight]{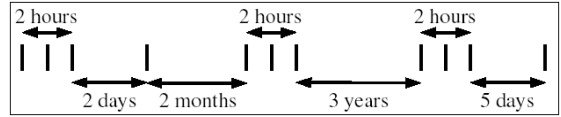}
\end{center}
\caption{This diagram shows the observing strategy for one
field (each vertical line is for one exposure,
exposure time depends on the filter, but is
typically of the order of 100 second)
}
\end{figure}

These characteristics, combined with the
excellent climatic conditions at the site,
provide images with a very good
quality.
Moreover, the images are pre-processed in quasi-real 
time with a pipeline called Elixir\footnote{See http://www.cfht.hawaii.edu/Instruments/Elixir/}, which flattens and
defringes each CCD and computes gross astrometry and photometry.

\subsection{The Legacy Survey}
This 5 years' project has started in march 2003, and is based on 3
different observing strategies:
\begin{itemize}
\item[-] The Wide Synoptic Survey covers 170 square degrees with all the 5
Megacam filters (u* g' r' i' z') down to approximatively i'=25.5. The main goal of
this survey is to study large scale structures and matter distribution in
the universe.
\item[-] The Deep Synoptic Survey covers only 4 square degrees, but down
to i'=28.4 and through the whole filter set. Aimed mainly at the
detection of 2000 type I Sne and the study of galaxy distribution, this
survey will allow an accurate determination of cosmological
parameters.
\item[-] The Very Wide Survey covers 1200 square degrees down to i'=23.5,
but only through 3 filters (g' r' i'). Initially conceived to discover and
follow 2000 Kuiper Belt Objects, we'll show that this observing
strategy can be used to detect optical afterglows.
\end{itemize}

\subsection{The Very Wide Survey}
Each field is observed several times, according to the strategy explained in Figure 1. This recurrence can be used to compare images
between them in order to detect variable or transient objects, such as GRB
afterglows

\section{OSGA Pipeline}
\subsection{Night Process}
The first part of the process consists in the reduction the useful information from 
700 Mo (the size of one Megacam image) to a few tens Mo. The pipeline automatically 
checks the presence of a new image as soon as it has been pre-processed by Elixir 
and starts the following treatment for each CCD:
\begin{itemize}
\item[-] Conversion of FITS image to JPEG format for future display on a web page, extraction of the FITS header and
addition of new entries, like magnitude and mu-max completness.
\item[-] Creation of a catalog of objects using SExtractor; in this catalog are mentioned 
X and Y position, RA and DEC, magnitude and mu-max, FWHM and a SExtractor flag
\item[-] Astrometric calibration using USNO catalog
\item[-] Sorting of objects according to their astrophysical properties into stars, 
galaxies, faint objects, saturated objects and cosmic rays
\end{itemize}

The processing of one image doesn't last more than 3 minutes.
All these results are then summarized in real time on an automatically generated HTML 
web page and can be viewed by everyone on the Internet. Using an interactive script, 
collaboration members are able to directly validate or not the night process, and so
allow the second stage of the processing which involves the comparison of the image
with previous images of the same field.

\begin{figure}
\begin{center}
\includegraphics[height=.4\textheight]{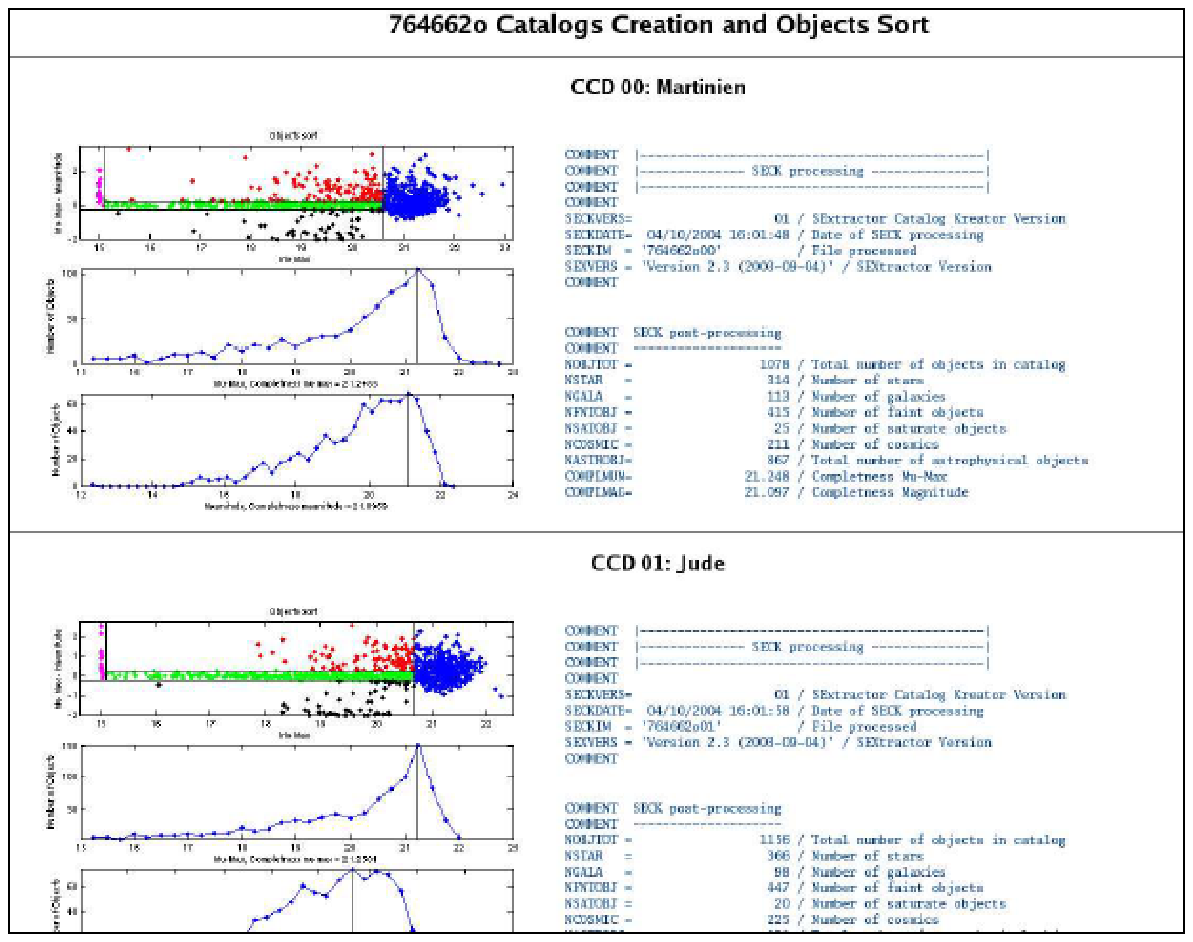}
\end{center}
\caption{This is a snapshot of one of the "catalog creation" web page. For each
CCD, 3 graphs show, respectively, the object sorting (where saturated objects
are in the leftmost part and faint object in the rightmost, whereas in the
central part, the box defines from top to bottom galaxies, stars and cosmic
rays), and the mu-max and magnitude completeness. The right panel show some
information added to the header, like the number of objects for each category
and the value of the mu-max and magnitude completeness.
}
\end{figure}

\subsection{Day Process}
At the end of each night, the second part of the process is launched. Its goal is first to list all the possible comparisons
between images just taken. To be possible, a comparison must involve images of the same field, with 
similar exposure times and the same filter. In a second time, the process checks if the field has already been observed during this run
(a run lasts about 3 weeks), and in this case starts the comparison between the two best quality images of each
night. The comparison process can be broken down as follow:
\begin{itemize}
\item[-] Classification of astrophysical objects into three categories: unchanged, luminosity variable and alone. Unchanged
objects are used for photometric calibration
\item[-] Research of asteroids and objects having significant variation in luminosity. Those objects are then extracted and stored in a
special catalog
\item[-] Generation of a few graphs and cut-out of thumbnails around the positions of interesting objects
\end{itemize}
The comparison of two or three images takes less than 5 minutes. Only a few ten objects are detected as variable in a whole comparison, whereas
the catalogs contain more than one hundred thousand objects.
All the information about interesting variable objects and asteroids is gathered on a self-generated interactive HTML
web page. Then a member of the collaboration has to reject false detections (like cosmic rays or CCD defects for instance)
or validate objects which are trully variable.

\begin{figure}
\begin{center}
\includegraphics{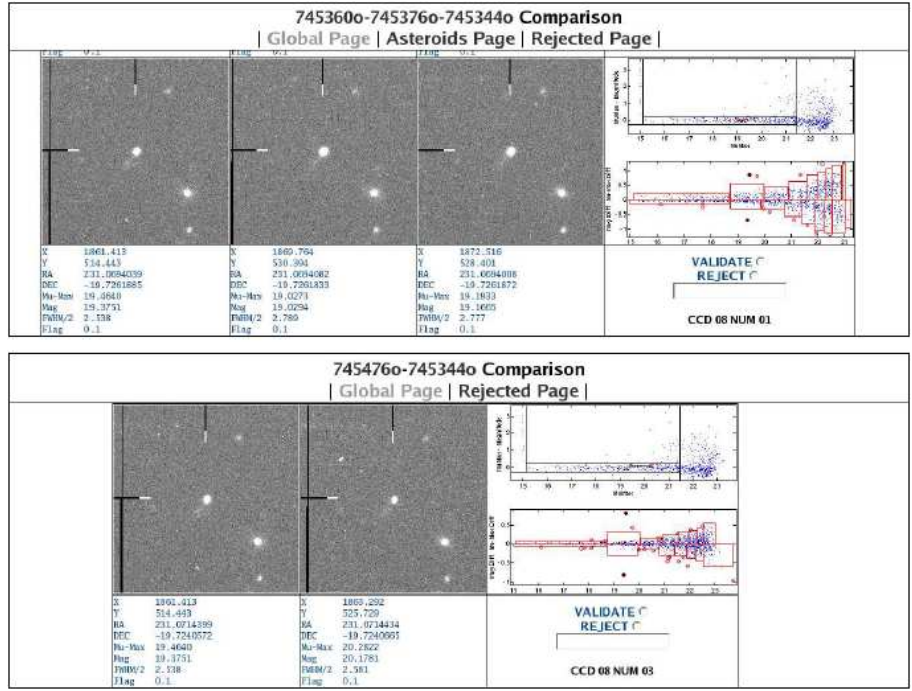}
\end{center}
\caption{A very interesting object can be seen on these 2 snapshots. Not only it has been detected as variable in a one night
triple comparison but also in a double comparison with a image taken a few days before. Although we cannot conclude
about the nature of this transient source, this detection shows us that the whole process works correctly.
The object is not a GRB afterglow as it appears in the DSS.
}
\end{figure}

\section{Conclusions}
We have shown that optical afterglow detection is possible using Megacam images from the Very wide survey at the CFHT and a real time comparison pipeline.
A few afterglows per year can reasonably be expected. 
The real time process has started in November 2004; though a few interesting
objects have been detecting, no GRB afterglow has been discovered yet.
Additional informations on this search can be found on the web page of the RTAS http://www.cfht.hawaii.edu/\~{}grb/

\end{document}